\documentclass[prl,twocolumn,showpacs,amsmath,amssymb,superscriptaddress]{revtex4}

\usepackage{graphicx}%Include figure files
\usepackage{dcolumn}%Align table columns on decimal point
\usepackage{bm}% bold math
\usepackage{epsfig}

\newcommand{\vectg}[1]{\mbox{\boldmath ${#1}$}}

\begin{document}

\title{Can Nonequilibrium Spin Hall Accumulation be Induced in Ballistic Nanostructures?}

\author{Branislav K. Nikoli\' c}
\affiliation{Department of Physics and Astronomy, University
of Delaware, Newark, DE 19716-2570, USA}
\author{Satofumi Souma}
\affiliation{Department of Physics and Astronomy, University
of Delaware, Newark, DE 19716-2570, USA}
\author{Liviu P. Z\^ arbo}
\affiliation{Department of Physics and Astronomy, University
of Delaware, Newark, DE 19716-2570, USA}
\author{Jairo Sinova}
\affiliation{Department of Physics, Texas A\&M University, College Station, TX 77843-4242, USA}

\begin{abstract} We demonstrate that flow of longitudinal unpolarized current through
a {\em ballistic}  two-dimensional electron gas with Rashba spin-orbit coupling will
induce nonequilibrium spin accumulation which has {\em opposite} sign for the two
lateral edges and it is, therefore, the principal observable signature  of the spin Hall
effect in two-probe semiconductor nanostructures. The magnitude of its out-of-plane
component is gradually diminished by static disorder, while it can be enhanced  by an
in-plane transverse magnetic field. Moreover, our prediction of the longitudinal component
of the spin Hall accumulation, which is insensitive to the reversal of the bias voltage,
offers a {\em smoking gun} to differentiate experimentally between the extrinsic, intrinsic,
and mesoscopic spin Hall mechanisms.
\end{abstract}

\maketitle

{\em Introduction}---When electric current flows along a conductor subjected
to a perpendicular magnetic field, the Lorenz force deflects the charge carriers
creating a transverse Hall voltage between the lateral edges of the sample. The
normal Hall effect is one of the most familiar phenomena, as well as a widely utilized
tool, in condensed matter  physics~\cite{classical_hall}. In the absence of external magnetic
field, more esoteric Hall-type effects involving electron spin become possible in
paramagnetic systems with spin-orbit (SO) couplings---the opposite spins can be separated
and then accumulated  on the lateral edges when they are transported by a {\em pure} (i.e., not
accompanied by any net charge current) spin Hall current flowing in the transverse direction
in response to unpolarized charge current in the longitudinal direction. For
instance, the SO dependent scattering off impurities, which  deflects spin-$\uparrow$
and spin-$\downarrow$ electrons of an unpolarized beam in opposite directions, and it is partially
responsible for the anomalous Hall effect in ferromagnetic metals~\cite{classical_hall}, has
been invoked in early studies to predict the {\em extrinsic} (i.e., due to impurity scattering)
spin Hall effect~\cite{extrinsic}.

\begin{figure}
\centerline{\psfig{file=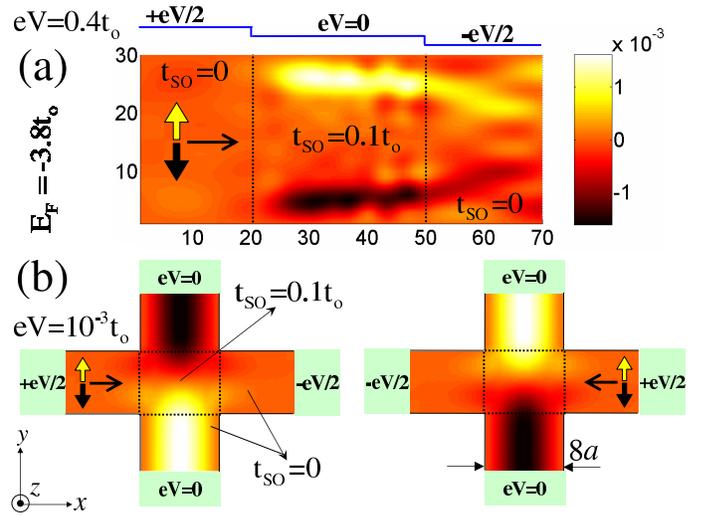,scale=0.36,angle=-90}} 
\caption{(color online). (a) The out-of-plane component $\left< S_z ({\bf r}) \right>$  of the nonequilibrium
spin accumulation induced by  nonlinear quantum transport  of unpolarized
charge current injected from the left lead into a two-terminal {\em clean} 2DEG (of size $L=30a > L_{\rm SO}$, $a \simeq 3$ nm) nanostructure with the Rashba SO coupling $t_{\rm SO}=0.1t_{\rm o}$ and spin precession length $L_{\rm SO} \approx 15.7a$.  Panel (b) shows how lateral spin-$\uparrow$ and spin-$\downarrow$ densities will {\bf flow} in opposite directions through the attached transverse ideal  ($t_{\rm SO} = 0$) leads to generate a linear response spin Hall current $[I_y^s]^z$ out of four-terminal 2DEG ($L=8a < L_{\rm SO}$) nanostructures~\cite{nikolic}, which changes  sign $[I_y^s]^z(-V)=-[I_y^s]^z(V)$ upon reversing the bias voltage.}\label{fig:accumulation_2d}
\end{figure}

The pursuit of all-electrical spin current induction and manipulation in
semiconductor spintronics~\cite{spintronics} has rekindled interest in the realm
of the spin Hall effect(s). In particular, recent theoretical arguments have unearthed
possibility for pure transverse spin Hall current that is several order of  magnitude
greater  than in the case of the extrinsic effect, arising  due to {\em intrinsic}
mechanisms related to the spin-split band structure in SO coupled bulk~\cite{sinova,murakami}
or mesoscopic~\cite{nikolic,sheng,ewelina} semiconductor systems. While these theories are
formulated in terms of not directly observable spin currents (which are the Fermi-sea 
quantity and not conserved in infinite homogeneous systems of the intrinsic effect~\cite{sinova,murakami}, 
or the Fermi-surface property and conserved ones in the transverse electrodes attached to 
mesoscopic samples~\cite{nikolic,sheng,ewelina}), their detection requires to measure 
nonequilibrium spin accumulation that they would deposit at the sample edges~\cite{extrinsic,kato,wunderlich}. 
The controversy in theoretical interpretations~\cite{zhang,bernevig} of recent breakthrough observations 
of the spin Hall accumulation~\cite{kato,wunderlich} is largely due to the fact that  {\em no theory exists} 
that demonstrates the existence of such accumulation in {\em ballistic}  SO coupled finite-size structures accessible to experiments~\cite{kato,wunderlich}.

\begin{figure}
\includegraphics[scale=1.0]{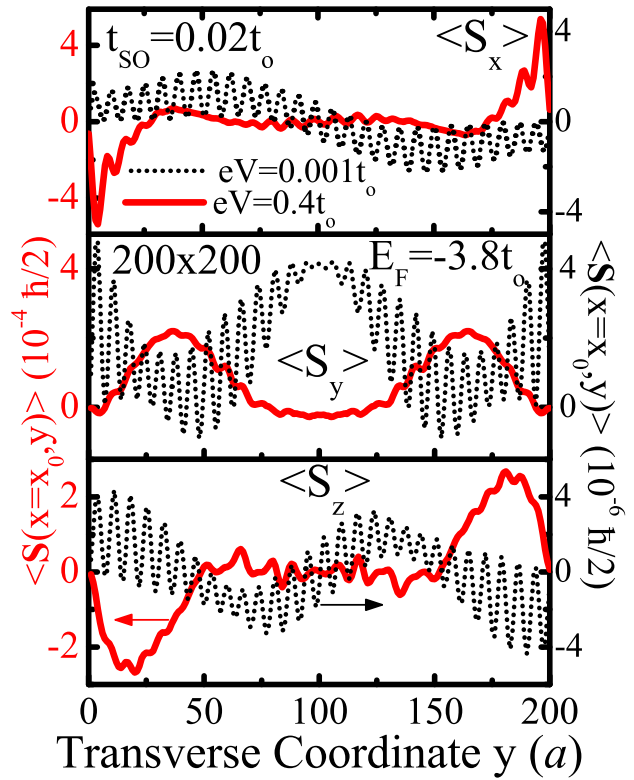}
\caption{(color online). The one-dimensional transverse spatial profile of the spin accumulation
$\left< {\bf S}(x=78a,y) \right>$ across the $200a \times 200a$ 2DEG with the Rashba 
SO coupling $t_{\rm SO}=0.02t_{\rm o}$ through which {\em ballistic} quantum transport takes place in the nonlinear regime $eV=0.4t_{\rm o}$  (solid lines) or the linear regime $eV=10^{-3}t_{\rm o}$ (dotted lines). The 
width of the  edge peaks of $\left< S_z(x=78a,y) \right>$ is $\approx L_{\rm SO}/2=\pi a t_{\rm o} /4t_{\rm SO}$.
 }\label{fig:nonlinear}
\end{figure}

Thus, to answer the question posed in the title of this Letter, we formulate a Landauer-Keldysh approach~\cite{keldysh,caroli} to the spin Hall accumulation problem in two-terminal devices, 
which treats phase-coherent spin-charge transport while taking into account all boundaries, 
interfaces, and electrodes of the semiconductor nanostructure~\cite{landauer}. We predict a nonequilibrium spin accumulation induced at the lateral edges of a ballistic two-dimensional electron gas (2DEG) with Rashba SO coupling attached to two ideal (interaction-free) semi-infinite leads as a response to unpolarized charge current flowing through the sample, as shown in Fig.~\ref{fig:accumulation_2d}(a). 
The electric field along the $z$-axis, which confines electrons within the $xy$-plane of quantum well in semiconductor heterostructure, generates an effective momentum-dependent magnetic field ${\bf B}_{\rm R}({\bf p})$  [which does not break time-reversal invariance] due to the Rashba SO coupling~\cite{spintronics,rashba_eq}. The  pattern of local spin density throughout the device is obtained from  
\begin{equation} \label{eq:accumulation}
\left< {\bf S}({\bf r}) \right> = \frac{\hbar}{2} \int\limits_{E_F - eV/2}^{E_F + eV/2} \frac{dE}{2  \pi i} {\rm Tr}_{\rm spin} \, [\hat{\bm \sigma} {\bf G}^<({\bf r},{\bf r};E, V)],
\end{equation}
where the exact spin-dependent lesser nonequilibrium Green function ${\bf G}^<({\bf r},{\bf r}^\prime;E,V)$~\cite{keldysh} is evaluated in the {\em steady-state} quantum transport  through the semiconductor nanostructure attached to external probes~\cite{caroli}. Here $V$ is the applied  bias voltage between the  leads and $\frac{\hbar}{2} \hat{\bm \sigma}$ is the spin-$\frac{1}{2}$ operator. In the 
nonlinear  phase-coherent transport regime, Fig.~\ref{fig:accumulation_2d}(a) and Fig.~\ref{fig:nonlinear} demonstrate  that transverse profile of the out-of-plane  $\left< S_z(\bf r) \right>$ component of the spin accumulation develops two peaks  of opposite sign at the lateral edges of the 2DEG. Upon reversing the bias voltage, the edge peaks flip their sign $\left< S_z({\bf r}) \right>_{-V} = - \left< S_z({\bf r}) \right>_{V}$. Around the left-lead/2DEG interface, $\left<S_z({\bf r}) \right> \rightarrow 0$ in Fig.~\ref{fig:accumulation_2d}(a) is suppressed since unpolarized electrons are injected at this contact.

All three features of $\left< S_z(\bf r) \right>$, delineated here for phase-coherent transport regime at low-temperatures, are remarkably similar to the general phenomenology
of the spin Hall effect demonstrated convincingly in two very recent experiments~\cite{kato,wunderlich}. 
These experiments focus on the optical detection of the spin Hall accumulation of opposite sign on the lateral edges of two-probe semiconductor structures (in the semiclassical transport regime at finite-temperatures): (i) weakly SO coupled 3D films of GaAs or strained InGaAs~\cite{kato}; and (ii) strongly SO coupled 2D hole gases~\cite{wunderlich}. The former experiment has been interpreted  as the manifestation of the extrinsic effect~\cite{extrinsic,zhang} due to small resulting polarization ($\sim 0.03$\%) and no SO splitting of the band structure~\cite{kato}, while
the later is consistent with the intrinsic mechanisms~\cite{sinova,bernevig} because of 
spin-split quasiparticle energies and much larger spin polarization ($\sim 1$\%)~\cite{wunderlich}.

However, it has been estimated that spin Hall accumulation 
induced by the extrinsic effect is far below present experimental
sensitivity~\cite{bernevig}. On the other hand, it has been
argued~\cite{zhang, rashba_eq} that the intrinsic spin Hall
current in the bulk, determined only by the equilibrium
distribution function and spin-split Bloch band structure~\cite{sinova,murakami}, 
does not really transport spins, so that no spin accumulation at the edges is 
possible in the absence  of impurities~\cite{zhang}. We bring such debates to an end in 
Fig.~\ref{fig:accumulation_2d}(b) by attaching two additional transverse ideal semi-infinite 
leads at the lateral edges of a perfectly clean 2DEG and showing how spin-$\uparrow$ and  spin-$\downarrow$ 
densities will {\em flow} through those leads in {\em opposite}  transverse 
directions to generate the spin Hall current driven by mechanisms~\cite{spin_force} on the 
{\em mesoscopic} scale~\cite{wees} set by the spin precession length~\cite{nikolic}.  
Thus, Fig.~\ref{fig:accumulation_2d}(b) demonstrates convincingly that spin 
Hall effect in SO coupled ballistic nanostructures can be used as an all-electrical 
semiconductor-based spin injector~\cite{spintronics}.

The attempts~\cite{shytov,ma} to understand the spin Hall accumulation in macroscopic {\em disordered} 
Rashba spin-split 2DEG attached to two massive electrodes have reached contradictory conclusions 
(vanishing $\left< S_z(\bf r) \right>$ on the lateral edges~\cite{shytov} vs. $\left< S_z(\bf r) \right> \neq 0$ at the edges and within the sample~\cite{ma}), which can be traced to an incomplete treatment of experimentally relevant measuring geometry within the semiclassical diffusion equation approaches. A plethora 
of spin transport phenomena illustrate the need to treat the whole device geometry due to the presence of SO couplings, even in the semiclassical transport regime~\cite{wees}. For example, spin relaxation in confined 
Rashba  structures is quite different from the bulk D'yakonov-Perel' (DP) spin
relaxation~\cite{spintronics} because of the transverse
confinement effects~\cite{purity} or chaotic vs. integrable
boundary scattering~\cite{spin_ballistic}. Also, ${\bf B}_{\rm
R}({\bf p})$ in Rashba SO coupled wires is almost parallel to the
transverse direction~\cite{spin_force} (in contrast to the
infinite 2DEG where no unique spin quantization axis
exists~\cite{sinova}). Mesoscopic transport techniques,
developed to treat the whole measuring geometry as demanded by 
quantum coherence effects~\cite{landauer}, are well-suited to handle 
all relevant details of the spin Hall transport measurement 
setups~\cite{kato,wunderlich}. Therefore, we apply~\cite{caroli} the Keldysh Green 
functions~\cite{keldysh} to the  Landauer-type geometry~\cite{landauer} where finite-size sample 
is attached to two macroscopic reservoirs via  semi-infinite ideal leads (which
simplify the boundary conditions)---here the current is limited by
quantum transmission through a potential profile while power is
dissipated non-locally in the reservoirs~\cite{caroli,landauer}.

{\em Landauer-Keldysh Green function approach to nonequilibrium spin accumulation.}---The effective
mass Hamiltonian modeling the ballistic finite-size 2DEG with Rashba SO coupling in Fig.~\ref{fig:accumulation_2d} is given by~\cite{spintronics,rashba_eq}
\begin{equation} \label{eq:rashba}
\hat{H} = \frac{\hat{p}_x^2 + \hat{p}_y^2}{2 m^*}  + \frac{\alpha}{\hbar}
\left( \hat{p}_y \hat{\sigma}_x  - \hat{p}_x  \hat{\sigma}_y  \right) + V_{\rm conf}(x,y),
\end{equation}
where $(\hat{\sigma}_x,\hat{\sigma}_y,\hat{\sigma}_z)$ is the vector of the Pauli matrices, $(\hat{p}_x,\hat{p}_y)$ is the momentum operator in 2D space, $\alpha$ is the strength of the
Rashba SO coupling~\cite{rashba_eq}, and  $V_{\rm conf}(x,y)$ is the transverse confining
potential. In order to evaluate the nonequilibrium Green functions for a sample of
arbitrary shape attached to the ideal  leads,
we follow Ref.~\cite{caroli} and employ the local orbital basis. In this
representation, the Rashba Hamiltonian is expressed as~\cite{purity}
\begin{eqnarray}\label{eq:tbh}
\hat{H}=\sum_{{\bf m}\sigma} \varepsilon_{\bf m} \hat{c}_{{\bf
m}\sigma}^\dag\hat{c}_{{\bf m}\sigma}+\sum_{{\bf
mm'}\sigma\sigma'} \hat{c}_{{\bf m}\sigma}^\dag t_{\bf
mm'}^{\sigma\sigma'}\hat{c}_{{\bf m'}\sigma'},
\end{eqnarray}
where hard wall boundary conditions account for confinement on the $L \times L$ lattice with
lattice spacing $a$ (typically $a \simeq 3$ nm~\cite{purity}).  Here $\hat{c}_{{\bf m}\sigma}^\dag$ ($\hat{c}_{{\bf m}\sigma}$) is the creation (annihilation)  operator of an electron at the site ${\bf m}=(m_x,m_y)$. The generalized nearest neighbor hopping $t^{\sigma\sigma'}_{\bf mm'}=({\bf t}_{\bf mm'})_{\sigma\sigma'}$ accounts for the Rashba coupling
\begin{eqnarray}\label{eq:hopping}
{\bf t}_{\bf mm'}=\left\{
\begin{array}{cc}
-t_{\rm o}{I}_s-it_{\rm SO}\hat{\sigma}_y &
({\bf m}={\bf m}'+{\bf e}_x)\\
-t_{\rm o}{I}_s+it_{\rm SO}\hat{\sigma}_x &  ({\bf m}={\bf m}'+{\bf e}_y)
\end{array}\right.,
\end{eqnarray}
through the SO hopping parameter $t_{\rm SO}=\alpha/2a$ ($I_s$ is the unit
$2 \times 2$ matrix in the spin space). The direct correspondence between the continuous Eq.~(\ref{eq:rashba}) and the lattice Hamiltonian Eq.~(\ref{eq:tbh}) is established by using  $t_{\rm o}=\hbar^2/2m^*a^2$ for the orbital hopping and by selecting the Fermi energy $E_F=-3.8t_{\rm o}$ close to the bottom of the band at $-4.0t_{\rm o}$ to ensure the parabolic energy-momentum dispersion.

At time $t^\prime = -\infty$ the 2DEG and the leads are not connected, while the left and the
right lead are in their own thermal equilibrium with the chemical potentials $\mu_L$ and $\mu_R$, respectively, where $\mu_L=\mu_R+eV$. The adiabatic switching of the hopping parameter connecting
the leads and the 2DEG generates time evolution of the density matrix of the structure~\cite{caroli}.
The spin accumulation is obtained as the nonequilibrium statistical average $\left< ... \right>$  (with respect to the density matrix at $t^\prime=0$~\cite{keldysh}) of the spin-density  operator
$\left< \hat{{\bf S}}_{\bf
m} \right>=\frac{\hbar}{2}\sum_{\sigma \sigma^\prime} \vectg{\sigma}_{\sigma \sigma^\prime} \left< \hat{c}^\dag_{{\bf
m}\sigma} \hat{c}_{{\bf m}\sigma^\prime} \right>$, which is expressed via the lesser Green function
$\left< \hat{c}^\dag_{{\bf m}\sigma}\hat{c}_{{\bf m}'\sigma^\prime} \right>  = \frac{\hbar}{i}G^<_{{\bf m}'{\bf m},\sigma^\prime \sigma}(\tau=0) = \frac{1}{2\pi i}\int_{-\infty}^{\infty}dE G^<_{{\bf m}'{\bf m},\sigma^\prime \sigma}(E)$ thereby yielding Eq.~(\ref{eq:accumulation}). Here we utilize the fact that the two-time correlation  function  $G^<_{{\bf m}{\bf m}',\sigma\sigma'}(t,t')\equiv\frac{i}{\hbar}\left<\hat{c}^\dag_{{\bf
m}'\sigma'}(t')\hat{c}_{{\bf m}\sigma}(t)\right>$ depends only on $\tau=t-t^\prime$ in
stationary situations, so it can be Fourier transformed to energy $G^<_{{\bf m}{\bf m}',\sigma\sigma'}(\tau) = \frac{1}{2\pi\hbar}\int_{-\infty}^{\infty}dE G^<_{{\bf m}{\bf m}',\sigma\sigma'}(E)e^{iE\tau/\hbar}$. The matrix ${\bf G}^<(E)$ is obtained from the Keldysh equation ${\bf G}^<(E) = {\bf G}^r(E) {\bm \Sigma^<}(E){\bf G}^a(E)$. Within the effective single-particle picture, this equation can be solved exactly in the noninteracting electron approximation by evaluating the retarded ${\bf G}^r(E)=[E-{\bf H}-U_{\bf m}-{\bm \Sigma}_L^r - {\bm \Sigma}_R^r]^{-1}$ and the advanced ${\bf G}^a(E)=[{\bf G}^r(E)]^\dag$ Green function matrices. In the absence of  inelastic scattering, the retarded self-energies ${\bm \Sigma}_L^r(E-eV/2)$ and ${\bm \Sigma}_R^r(E+eV/2)$ introduced  by the left and the right lead~\cite{caroli}, respectively, determine ${\bm \Sigma}^<(E)= -2i  [{\rm Im}\, {\bm \Sigma}_L(E-eV/2) f_L(E-eV/2) + {\rm Im}\, {\bm \Sigma}_R(E+eV/2) f_R(E+eV/2)]$  [$f_{L,R}$ are the Fermi distribution functions of the electrons injected from the macroscopic reservoirs  through the leads and ${\rm Im} \, {\bm \Sigma}_{L,R} = ({\bm \Sigma}_{L,R}^r -{\bm \Sigma}_{L,R}^a)/2i$]. The gauge invariance of measurable quantities with respect to the shift of electric potential by a constant is satisfied on the proviso that ${\bm \Sigma}_L^r$,  ${\bm \Sigma}_R^r$ depend explicitly on  the applied  bias voltage $V$ while ${\bf G}^r(E)$ has to include the electric potential landscape $U_{\bf m}$ within the sample~\cite{christensen}.

{\em Non-equilibrium spin accumulation in clean mesoscopic two-probe  2DEG structures.}---The
one-dimensional spatial transverse profile~\cite{kato} of $\left< {\bf S}(x=x_0,y) \right>$
across the macroscopic ($200a \times 200a$) 2DEG sample supporting steady-state charge current at zero-temperature  is plotted in Fig.~\ref{fig:nonlinear} for the {\em nonlinear} $eV \sim E_F$ as well as the
{\em linear} $eV \ll E_F$ transport regime. In the linear phase-coherent transport regime at low
temperatures only states at the  Fermi energy contribute to  $\left< {\bf S}({\bf r}) \right>$
in Eq.~(\ref{eq:accumulation}), which  therefore encodes the  information about their wave
function and oscillates within the sample. Figure~\ref{fig:nonlinear} also shows that the
width of the edge peaks of $\left< S_z({\bf r}) \right>$ is determined by the
spin precession length $L_{\rm SO}= \pi a t_{\rm o}/2t_{\rm SO}$~\cite{purity} (on which 
spin precesses by an angle $\pi$; $L_{\rm SO}$ also plays the role of the DP spin relaxation length~\cite{spintronics,spin_ballistic} in weakly disordered systems) in our two-probe device which 
is in the regime $L > L_{\rm SO} \gg \lambda_F$.

While experiments~\cite{wunderlich,kato} have searched for the out-of-plane spin accumulation
exhibiting opposite signs on the lateral edges as the evidence for the spin Hall effect, we predict
that {\em both} $\left< S_z({\bf r}) \right>$ and $\left< S_x({\bf r}) \right>$ satisfy this criterion.
The emergence of $\left< S_x({\bf r}) \right> \neq 0$ is the hallmark of the mesoscopic spin Hall effect~\cite{nikolic} in Rashba wires where Hamiltonian Eq.~(\ref{eq:rashba}) gives rise to the SO force operator~\cite{spin_force}  $\hat{\bf F} = 2 \alpha^2 m^*  (\hat{\bf p} \times {\bf z}) \otimes \hat{\sigma}_z/\hbar^3  - d V_{\rm conf}(\hat{y}){\bf y}/d \hat{y}$ which deflects spin-$\uparrow$ 
and spin-$\downarrow$ electronic wave packets in opposite transverse  directions,  while spin is at the same 
time precessing since injected $|\!\! \uparrow \rangle$, $|\!\! \downarrow \rangle$ (the eigenstates 
of $\hat{\sigma}_z$) are not the eigenstates of the Zeeman term  $\hat{\bm \sigma} \cdot {\bf B}_{\rm R}({\bf p})$ whose ${\bf B}_{\rm R}({\bf p})$-field is almost parallel to the $y$-axis~\cite{purity,spin_force}. 
This semiclassical picture~\cite{spin_force} heuristically explains the symmetry  properties of the quantum  transport induced accumulations in Fig.~\ref{fig:nonlinear} with respect to the bias voltage reversal, $\left< S_z({\bf r}) \right>_{-V} = - \left< S_z({\bf r}) \right>_{V}$ vs. $\left< S_x({\bf r}) \right>_{-V} = \left< S_x({\bf r}) \right>_{V}$, which can be tested experimentally~\cite{wunderlich}. Also, the $\alpha^2$-dependent transverse SO ``force'' accounts for the difference $\left< S_z({\bf r}) \right>_{-\alpha} = \left< S_z({\bf r}) \right>_{\alpha}$ vs. $\left< S_x({\bf r}) \right>_{-\alpha} =- \left< S_x({\bf r}) \right>_{\alpha}$. Finally, Fig.~\ref{fig:nonlinear}(a) confirms that $\left< S_z({\bf r}) \right>$ spin densities will move away from the lateral edges upon entering the right lead where $\alpha \equiv 0$ and the transverse SO force deflecting the spins is absent.

In contrast to the out-of-plane $\left< S_z ({\bf r}) \right >$ and the in-plane longitudinal $\left< S_x ({\bf r}) \right >$ spin Hall accumulation, the transverse in-plane spin accumulation $\left< S_y ({\bf r}) \right>$  has the same sign on both lateral edges. Thus, it cannot originate from the spin Hall effect. Instead, it is an analog of the magneto-electric effect---where electric field induces spin polarization of conduction electrons and nonequilibrium magnetization ${\bf M} = \beta {\bf E}$---which was argued to occur in {\em disordered} paramagnetic systems (metals with SO scattering off impurities~\cite{levitov} or diffusive Rashba spin-split 2DEG~\cite{edelstein,shytov}), and has finally been observed experimentally~\cite{ganichev}. It arises due to the  combined effect of SO coupling, inversion asymmetry, and time-reversal symmetry breaking by electron scattering (which is necessary due to ${\bf M}$ being $t$-odd while ${\bf E}$ is $t$-even). However, here we predict that passage of electric current through {\em ballistic} sample (${\bf E}=0$ inside 2DEG) will also induce the transverse in-plane magnetization, where finite quantum point contact conductance of
the nanostructure in Fig.~\ref{fig:accumulation_2d} (i.e., the corresponding dissipation in the  reservoirs~\cite{landauer,caroli}) signifies the time-reversal symmetry breaking.

The key issue for spintronics applications is to detect (at least
indirectly~\cite{extrinsic,ewelina}) the pure spin Hall current
flowing out of  multiterminal structures. When transverse leads
(e.g., labeled by 2 and 3) are attached to the lateral edges of
2DEG, the nonequilibrium spin accumulation will push the pure
spin current $\frac{\hbar}{2e}(I_2^\uparrow - I_2^\downarrow)$
into them, as shown in Fig.~(\ref{fig:accumulation_2d})(b). 
The properties of its spin Hall $[I_2^s]^z = - [I_3^s]^z$, $[I_2^s]^x = -
[I_3^s]^x$ and spin ``polarization'' $[I_2^s]^y = [I_3^s]^y$
components predicted in Ref.~\cite{nikolic} are exactly the
same as can be inferred from the profiles of $\left< {\bf S} ({\bf r}) \right >$ in Fig.~\ref{fig:nonlinear}.

\begin{figure}
\includegraphics[scale=0.8]{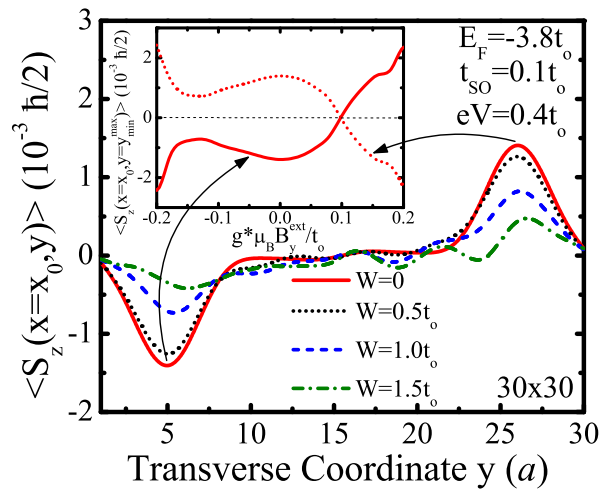} 
\caption{(color online). The transverse profile $\left< S_z(x=13a,y) \right>$ of spin Hall 
accumulation as a function of spin-independent disorder $W$ setting the mean free path $\ell \approx 21.5a/W^2$ [the ballistic limit $W=0$ is the transverse profile of $\left< S_z({\bf r}) \right>$ in  Fig.~\ref{fig:accumulation_2d}(a)]. The inset shows the effect of the external in-plane magnetic
field $g^* \mu_B B_y^{\rm ext}$ on $\left< S_z(x=13a,y=y^{\rm max}_{\rm min}) \right>$ at its maximum $y^{\rm max}=26a$ or minimum $y_{\rm min}=5a$ peak values.}\label{fig:disorder}
\end{figure}

{\em The effect of disorder and in-plane magnetic field on the mesoscopic spin Hall accumulation.}---We 
show in Fig.~\ref{fig:disorder} that disorder, introduced as the random potential $\varepsilon_{\bf m} \in [-W/2,W/2]$ in Eq.~(\ref{eq:tbh}), is gradually diminishing the amplitude of the original clean limit edge peaks of $\left< S_z ({\bf r}) \right>$ while leaving their position {\em intact}. Thus, the magnitude of the peaks alone does not provide single criterion to differentiate between possible underlying  mechanisms of the spin Hall effects~\cite{kato,bernevig}. We offer here yet another experimental tool (in addition to $\left< S_x({\bf r}) \right>_{-V} = \left< S_x({\bf r}) \right>_{V}$ test) to identify the mesoscopic spin Hall mechanism by demonstrating in the inset of Fig.~\ref{fig:disorder} that an external in-plane transverse  magnetic field $B_y^{\rm ext}$, which in strictly 2D systems does not affect orbital degrees of freedom,  will not destroy the edge peaks of $\left< S_z ({\bf r}) \right>$, except when the Rashba field ${\bf B}_{\rm R}({\bf p})$  term is cancelled by the Zeeman spin splitting term $g^* \mu_B B_y^{\rm ext}$.

\begin{acknowledgments}
We are grateful to E. I. Rashba, L. Sheng, A. H. MacDonald, and T. Jungwirth for enlightening discussions.
This research was supported in part by ACS grant PRF-41331-G10.
\end{acknowledgments}

%********************references************************************************************************

%*****************************************************************

\end{document}